\documentclass[10pt,conference,switch]{IEEEtran}
\IEEEoverridecommandlockouts
\usepackage{amsmath,amssymb,amsfonts}
\usepackage{algorithmic}
\usepackage{graphicx}
\usepackage{textcomp}
\usepackage[table]{xcolor}
\def\BibTeX{{\rm B\kern-.05em{\sc i\kern-.025em b}\kern-.08em
    T\kern-.1667em\lower.7ex\hbox{E}\kern-.125emX}}

\usepackage{enumitem}

\usepackage{todonotes}
\usepackage[numbers, sort&compress]{natbib}
\usepackage{url}
\usepackage{booktabs}
\usepackage{caption}
\usepackage{multirow}
\usepackage{censor}
\usepackage{hyperref}
\usepackage{soul}

\usepackage{subcaption}
\usepackage{mathrsfs}
\usepackage{footmisc}

\usepackage{todonotes}

\newcommand{\email}[1]{\href{mailto:#1}{#1}}

\begin{document}

\title{The (ab)use of Open Source Code to Train Large Language Models}

\author{
\IEEEauthorblockN{Ali Al-Kaswan}
\IEEEauthorblockA{\textit{Delft University of Technology} \\
Delft, The Netherlands \\
\email{a.al-kaswan@tudelft.nl}} 
\and
\IEEEauthorblockN{Maliheh Izadi}
\IEEEauthorblockA{\textit{Delft University of Technology} \\
Delft, The Netherlands \\
\email{m.izadi@tudelft.nl}}
}

\maketitle

\begin{abstract}
In recent years, 
Large Language Models (LLMs) have gained significant popularity 
due to their ability to generate human-like text 
and their potential applications in various fields, 
such as Software Engineering.
LLMs for Code are commonly trained 
on large unsanitized corpora of source code scraped from the Internet. 
The content of these datasets is memorized and emitted by the models, 
often in a verbatim manner. 
In this work, we will discuss the security, privacy, and licensing implications of memorization. 
We argue why the use of copyleft code to train LLMs is a legal and ethical dilemma. 
Finally, we provide four actionable recommendations to address this issue. 
\end{abstract}

\section{Language Models for Code}

Large Language Models (LLMs) have gained significant attention 
in the field of Natural Language Processing (NLP) 
in recent years due to their ability 
to perform a wide range of NLP tasks with impressive accuracy. 
These models, trained on massive amounts of data, improve in accuracy as they grow from millions to billions of parameters.
%
LLMs for code are trained on massive amounts of data and can learn the structure and syntax of programming languages, 
making them well-suited for tasks such as 
code summarization, generation, and completion~\cite{al2023extending,izadi2022codefill}. 
LLMs are even making their way into commercial products like GitHub's Copilot, Replits's GhostWriter and Tabnine.  
Meanwhile, some have identified that LLMs can memorize large swaths of training data ~\cite{carlini2021extracting}. 
Memorization enables the extraction of the data using Data Extraction Attacks. 
Some attacks have even been able to extract addresses and other personal information from public models~\cite{carlini2021extracting}. 
Memorization also impacts LLMs for code, with all its associated consequences. We will discuss these consequences in three categories: security, privacy and licensing.

\section{Security Implications}
Text memorization has strong security implications. 
Firstly, massively mined code datasets 
are not sanitized or manually curated, 
the datasets could therefore contain many biases,\footref{gpt2phone}
and instances of badly written or buggy and insecure code. 
A recent study found that around 40\% of GitHub Copilot's code generations 
for MITRE's top 25 Common Weakness Enumerations, a list of the most dangerous software weaknesses were found to be vulnerable~\cite{hammond2022asleep}. If these models become more prevalent and trusted, they can introduce more vulnerable code into software.

\section{Privacy Implications}

Memorization enables adversaries to access training data, and everything contained within, simply by accessing the model. This has major privacy implications since code can contain private information.  
Think of credentials, API keys, directory structures, logged info, or in-code discussions by developers. Code can also contain personal information like emails or contact information. 
If personal data is published on the Internet, the data could be retracted and deleted from the source. But once it is mined and used to train an LLM, the information is forever embedded in a compact representation, which is queryable at scale. With query access to these models, an adversary can potentially extract this data~\cite{carlini2021extracting} and
threaten Internet users' privacy. 
There are many reasons why one could publicly share private information; (1) simply by accident, or (2) a malicious actor could share this information in a doxing campaign~\cite{carlini2021extracting}. 
Even if the data is published willingly, the owner has a certain use and audience in mind and might not wish to share this information with the entire world. This is referred to as the re-purposed data problem.\footnote{Does GPT-2 Know Your Phone Number?: \url{http://archive.is/LxsyA} \label{gpt2phone}} 

\section{Licensing}
Publicly available source code is also subject to licences, some of which heavily regulate the use of the material.
Initially, developers raised concerns about licensed code on social media. GitHub Copilot could be prompted to produce verbatim copies of copyrighted code, without providing the required attribution or licence terms.\footnote{Matrix Transpose: \url{http://archive.is/YU5Bl}} Similarly, Copilot was producing copyrighted code while attributing the wrong author and providing the wrong license.\footnote{Fast Inverse Square: \url{http://archive.is/HNiyg}}
Later, a lawsuit was filed against GitHub, Microsoft and OpenAI, claiming that Copilot is violating the licence of open-source code.\footnote{GitHub Complaint (p.26): \url{http://archive.is/3PFAs}\label{GHComplaint}}


Broadly, open-source code is licensed under two types of licences.
\textbf{Permissive licenses}, allow users to use, modify, and distribute the software for any purpose, without requiring that the user share their work. 
\textbf{Non-permissive licenses}, also known as "copyleft" licenses, require that users freely share their own software under the same licence if they distribute the software or any \textit{derivatives} of it. 
Creating closed or commercial software based on non-permissively licensed code is unethical and possibly even illegal~\cite{sun2022coprotector}. But this does raise the following question:
\textbf{Does training LLMs on copyleft code infringe on their license?} 

Firstly, we must determine how many LLMs for code are trained on copyleft code. 
Looking at some of the most popular code models, we can observe that the vast majority are trained on open-source code. CodeBERT and CodeT5 are trained on CodeSearchNet, which contains copyleft code.
We also found that CodeBERT, CodeGen, and CodeClippy make use of The Pile, a collection of $22$ datasets, one of which is a GitHub dataset containing copyleft data.

We found that only InCoder makes an effort to prevent training on copyleft code. InCoder does however make use of a dataset of StackOverflow answers, which are licensed under varying CC-BY-SA licences, all of which require attribution.\footnote{StackOverflow license: \url{http://archive.is/obaoy}} 

Despite the public attention, it is not completely clear whether Codex, the model behind Copilot, is trained on non-permissive code. Many imply that it does,\footnote{Comment on Copilot and OSS: \url{http://archive.is/6gEOU}\label{eeveeCopilot}} citing the copying of copyleft code and the fact that the system has encountered a copy of the GPL licence many times during training. The training data for Codex is not publicly available, and neither OpenAI nor Github have provided any clarification.\footnote{If Software is My Copilot, Who Programmed My Software?: \href{http://archive.is/pilW2}{Link}\label{sfcCopilot}}

LLMs for code can be seen as derivatives of their training data. So unless the model is published under the same licence as the training data and includes the copyright notice, this would be a clear violation. Moreover, many licences are not inter-compatible, i.e., the inclusion of code licensed under them automatically warrants an infringement as the combined licence agreements contain irreconcilable conditions. 

Some opponents,\footnote{\label{RedaGithub}GitHub Copilot is not infringing your copyright: \url{http://archive.is/PYlm5}} including OpenAI, argue that the use of public code is an instance of transformative fair use, which is a defence that allows the use of copyrighted works in new and unexpected ways and exists in many jurisdictions including the US.\footref{RedaGithub} Yet, it is still unclear whether the fair use defence applies to ML-systems,\footref{GHComplaint} as it has not yet been tested in court. Furthermore, the fair use argument is sometimes based on the assumption that models do not memorize and emit training data, which is false. Even if the fair use argument protects the use of the data, the verbatim outputting might not be protected. 

A moral argument can also be made on this issue. Training LLMs on copyleft code goes against the will of some open-source developers, who share their code for the betterment of society and who believe in the principle of free and open software so profoundly, they're willing to add a full legal clause to their work to perpetuate this ideal. The use of their work, without attribution, especially by commercial parties, is not what they had in mind.

Finally, some researchers have also proposed a different approach to this issue by letting the authors of open-source code take matters into their own hands. Using data poisoning techniques, the authors can reduce the performance and embed watermarks into the models~\cite{sun2022coprotector}.

\section{Discussion and Recommendations}
To conclude we recommend the following:
\begin{itemize}
    \item The ML community should carefully consider the licence of their training material, from both a legal and an ethical point of view. The authors of published LLMs should be transparent about the licences of their training material.
    \item More research should be conducted on the nature and proportionality of text memorization in LLMs for code and LLMs in general. Other topics include memorized text extraction and prevention.
    \item Lawmakers should clarify whether the use of copyleft code (and copyrighted materials in general) and text to train LLMs constitutes fair use and under which conditions this clause applies.
    \item Finally, the software engineering community should clarify their stance on this issue. Developers could make informed decisions and clearly denote if their source code can be used to train AI models. LLMs for code are likely to stay and bring new tools that change the way software is engineered. So the community needs to answer important questions on this matter. For instance, should open-source code be allowed for training these models? If so, should the developers be credited and compensated, and under which license should the models be released? Alternatively, do we need to revise current code licenses to clarify the community's stance?\footnote{Additional Reading Material: \href{https://github.com/AISE-TUDelft/nlbse23_reading_list}{Link to our GitHub Repository}} 
\end{itemize}

\bibliographystyle{IEEEtranN}
\bibliography{references.bib}

\end{document}